\pdfoutput=1
\documentclass[12pt,eqsecnum]{article}
\usepackage{fancyhdr}
\usepackage{datetime} 

\usepackage{amsmath,amssymb,amsbsy,exscale}
\usepackage{graphicx}
\usepackage{epsfig}
\usepackage{multicol}
\usepackage{mathrsfs}
\usepackage{mathtools}
\usepackage{blindtext}
 \usepackage{fancyhdr}
\usepackage{hyperref}
\usepackage{cite}

\usepackage{enumerate}

\usepackage{dsfont}
\usepackage{appendix}

\usepackage{parskip}

\usepackage{xspace}
\usepackage{braket}
\usepackage{array}
\usepackage{feynmp}
\usepackage{graphicx,rotating,colortbl}

\usepackage{pinlabel}
\usepackage[top=1.2in, bottom=1.2in,left=0.8in, right=0.8in,includefoot]{geometry}

\definecolor{nicered}{rgb}{0.7,0.1,0.1}
\definecolor{nicegreen}{rgb}{0.1,0.5,0.1}

\definecolor{rosso}{cmyk}{0,1,1,0.4}
\definecolor{babypink}{rgb}{0.96, 0.76, 0.76}
\definecolor{babyblueeyes}{rgb}{0.63, 0.79, 0.95}
\definecolor{azure(colorwheel)}{rgb}{0.0, 0.5, 1.0}
\definecolor{amethyst}{rgb}{0.6, 0.4, 0.8}
\definecolor{MyDarkBlue}{rgb}{0,0.1,0.7}

\definecolor{secnum}{RGB}{13,151,225}
\definecolor{ptcbackground}{RGB}{212,237,252}
\definecolor{ptctitle}{RGB}{0,177,235}
\definecolor{blus}{cmyk}{1,1,0,0.1}
\definecolor{verdes}{cmyk}{0.99,0,0.59,0.65}
\definecolor{rossos}{cmyk}{0,1,1,0.55}
\definecolor{redy}{cmyk}{0,1,1,0.7}
\definecolor{greeny}{cmyk}{0.99,0,0.59,0.98}
\definecolor{green-go}{cmyk}{0.79,0,0.59,0.5}

\usepackage{hyperref, pdfsync, epstopdf, enumerate}
\hypersetup{colorlinks,bookmarksopen,bookmarksnumbered,citecolor={nicegreen},
linkcolor={MyDarkBlue},pdfstartview=FitH,urlcolor={amethyst}}
\usepackage{amsfonts}
\usepackage{slashed}

\usepackage{verbatim}
\usepackage{doi}
\usepackage{url}
\usepackage{array}

\usepackage{cite}
\usepackage{float}
\usepackage[labelfont=bf]{caption}
\usepackage{subcaption}
\usepackage{parskip}
\usepackage{hhline}

\numberwithin{equation}{section}

\usepackage{sectsty}

\usepackage{mdframed}
\usepackage{titletoc}

\titlecontents{lsection}
  [5.8em]{\sffamily}
  {\color{greeny}\contentslabel{2.3em}\normalcolor}{}
  {\titlerule*[1000pc]{.}\contentspage\\\hspace*{-5.8em}\vspace*{5pt}%
    \color{white}\rule{\dimexpr\textwidth-15.5pt\relax}{1pt}}

\usepackage{titlesec}
\titlelabel{\thetitle.\quad}

\newcommand{\tmtextbf}[1]{{\bfseries{#1}}}
\newcommand{\tmtextrm}[1]{{\rmfamily{#1}}}

\newcommand{\gappeq}{{\rlap{{\raise}.5ex\text{\ensuremath{>}}}{{\lower}.5ex\text{\ensuremath{\sim}}}}}
\newcommand{\lappeq}{{\rlap{{\raise}.5ex\text{\ensuremath{<}}}{{\lower}.5ex\text{\ensuremath{\sim}}}}}

\newcommand{\I}{\tmtextrm{1{\kern}-.24em l}}

\newcommand{\newc}{\newcommand}
\newc{\be}{\begin{equation}}
\newc{\ee}{\end{equation}}
\newc{\bal}{\begin{align}}
\newc{\eal}{\end{align}}
\newc{\ba}{\begin{eqnarray}}
\newc{\ea}{\end{eqnarray}}
\newc{\bea}{\begin{eqnarray*}}
\newc{\eea}{\end{eqnarray*}}
\newc{\D}{\partial}
\newc{\som}{\sin\omega}
\newc{\com}{\cos\omega}
\newc{\sth}{\sin\theta}
\newc{\cth}{\cos\theta}
\newc{\stom}{\sin^2\omega}
\newc{\ctom}{\cos^2\omega}
\newc{\stth}{\sin^2\theta}
\newc{\ctth}{\cos^2\theta}
\newc{\ie}{{\it i.e.} }
\newc{\eg}{{\it e.g.} }
\newc{\etc}{{\it etc.} }
\newc{\etal}{{\it et al.}}

\def\beq{\begin{equation}}
\def\eeq{\end{equation}}
\def\bea{\begin{eqnarray}}
\def\eea{\end{eqnarray}}

\let\eps=\epsilon

\let\alp=\alpha

\newcommand{\dd}{\text{d}}

\newcommand{\lapproxeq}{\lower .7ex\hbox{$\;\stackrel{\textstyle
<}{\sim}\;$}}
\newcommand{\gapproxeq}{\lower .7ex\hbox{$\;\stackrel{\textstyle
>}{\sim}\;$}}
\newcommand{\stackdown}[2]{\lower 1.4ex\hbox{$\;\stackrel{\textstyle{#1}}
{\scriptstyle{#2}}\;$}}

\makeatletter
\def\@xfootnote[#1]{%
  \protected@xdef\@thefnmark{#1}%
  \@footnotemark\@footnotetext}
\makeatother

\numberwithin{equation}{section}

\newc{\comm}[1]{}
\definecolor{NiceLightBlue}{RGB}{0, 0,255}
\definecolor{Goldenrod}{RGB}{218,165,32}
\definecolor{OrangeRed}{RGB}{255,69,0}
\definecolor{Lime}{RGB}{0,255,0}
\hypersetup{
colorlinks = true,
breaklinks,
urlcolor={NiceLightBlue}
,citecolor={NiceLightBlue}
,linkcolor={NiceLightBlue}
}
\newc{\tom}[1]{\textcolor{Lime}{#1}}

\begin{document}

\topmargin -1.0cm
\oddsidemargin -0.5cm
\evensidemargin -0.5cm

 
{\vspace{1cm}}
\begin{center}
\vspace{1cm}

 {\Large  \tmtextbf{ 
{Nonminimal Coleman--Weinberg Inflation with an $R^2$ term}
}} {\vspace{.5cm}}\\

\vspace{1.4cm}

{\large  {\bf Alexandros Karam\footnote[1]{{\href{mailto:alkaram@cc.uoi.gr}{alkaram@cc.uoi.gr}}}, Thomas Pappas\footnote[2]{{\href{mailto:thpap@cc.uoi.gr}{thpap@cc.uoi.gr}}} and Kyriakos Tamvakis\footnote[3]{{\href{mailto:tamvakis@uoi.gr}{tamvakis@uoi.gr}}}
}
\vspace{.3cm}

{\it }\

{\em  \normalsize 

~Department of Physics, University of Ioannina, GR--45110 Ioannina, Greece


}
\vspace{0.5cm}

}
\vspace{1.2cm}
 \end{center}
\noindent --------------------------------------------------------------------------------------------------------------------------------

\vspace{-0.3cm}

\begin{abstract}

\noindent {\normalsize We extend the Coleman--Weinberg inflationary model where a scalar field $\phi$ is non-minimally coupled to gravity with the addition of the $R^2$ term. We express the theory in terms of two scalar fields and going to the Einstein frame we employ the Gildener--Weinberg formalism, compute the one-loop effective potential and essentially reduce the problem to the case of single-field inflation. It turns out that there is only one free parameter, namely, the mixing angle between the scalars. For a wide range of this angle, we compute the inflationary observables which are in agreement with the latest experimental bounds. The effect of the $R^2$ term is that it lowers the value of the tensor-to-scalar ratio $r$.}

\end{abstract}

\vspace{.9cm}

\noindent --------------------------------------------------------------------------------------------------------------------------------





%

\newpage

\section{Introduction}

Most of the problems of the Big Bang cosmology can be solved if one postulates that the Universe underwent a quasi--de Sitter expansion in its early stages. This inflationary era, when treated quantum-mechanically, can also produce and amplify the small inhomogeneities which have resulted in the large scale structures and the anisotropy in the temperature of the cosmic microwave background (CMB)~\cite{Hawking1982, Starobinsky1982, Guth1982, Linde1983a} we observe today. 

One of the simplest and most successful inflationary models is that of Starobinsky~\cite{Starobinsky1980}. By extending the Einstein-Hilbert action with the addition of an $R^2$ term, one may express the inflationary observables, i.e. the scalar spectral index $n_s$ and tensor-to-scalar ratio $r$, in terms of the number of $e$-folds $N$ as~\cite{Mukhanov1981, Starobinsky1983}
\bea 
n_s &=& 1 - \frac{2}{N} \,, \\
r &=& \frac{12}{N^2}\,.
\eea
Then, for $N=55$ one obtains $n_s = 0.9636$ and $r = 0.004$. These values lie within the sweet spot of the latest results from the Planck collaboration~\cite{Akrami2018} which have recently further constrained these parameters to be
\bea 
n_s &=& 0.9649 \pm 0.0042 \quad \text{at} \quad 68 \% \,\, \text{CL} \,, \\
r_{0.002} &<& 0.064 \qquad \qquad \quad \,\,\, \text{at} \quad 95 \% \,\, \text{CL} \,.
\eea
Starobinsky inflation belongs to the general class of $f(R)$ models~\cite{Capozziello2006, Briscese2007, Nojiri2008, Nojiri2008a, DeFelice2010, Capozziello2011, Clifton2012, Rinaldi2014, Bamba2014, Broy2015a, Odintsov2016, Guzzetti2016} which are equivalent to the scalar-tensor theories of gravity~\cite{Wagoner1970, Damour1992, Damour1993, Barrow1995, Garcia-Bellido1995a, Boutaleb-Joutei1997, Boisseau2000, Morris2001, Esposito-Farese2001, Chiba2003, Clifton2012, Stabile2013, Chiba2013a, Obukhov2014, Jaerv2015a, Kuusk2016a, Vilson2017, Tambalo2017, Artymowski2017, Bhattacharya2017, Bhattacharya2017a, Burns2016, Karam2017, Karamitsos2018, Karamitsos2018a, Karam2018}. Other simple and popular scenarios have been the monomial potentials with integer powers of the inflaton field. These, however, are all now excluded with possibly the exception of linear inflation which gives $r = 0.066$~\cite{Artymowski2017, Kannike2016, Racioppi2017, Racioppi2018, Karam2018a}. Recently, popular ideas such as Higgs inflation have been combined with the Starobinsky model~\cite{Artymowski2015a, Bruck2015, Asaka2016, Artymowski2016, Kaneda2016, Calmet2016, Bruck2016a,   Wang2017, Ema2017, Mori2017, Pi2018, He2018, Gorbunov2018, Ghilencea2018, Wang2018a, Antoniadis2018, Gundhi2018}. The analysis of these models, however, becomes more complicated, since the appearance of the scalaron in the Einstein frame results in a two-field inflaton potential.

The fact that the $R^2$ term of the Starobinsky model dominates over the linear term during inflation suggests that at very high energies gravity is scale invariant. In recent years, numerous scale-invariant models have been proposed~\cite{Cooper1981, Shaposhnikov2009a, Garcia-Bellido2011, Khoze2013, Bezrukov2013a, Gabrielli2014, Salvio2014, Csaki2014a, Kannike2015, Barvinsky2015, Marzola2016a, Barrie2016, Marzola2016b, Rinaldi2016a, Farzinnia2016, Kannike2016, Karananas2016, Ferreira2016, Tambalo2017, Kannike2017, Artymowski2017, Racioppi2017, Ferreira2017, Salvio2017, Karam2017, Kannike2017a, Barnaveli2018, Ferreira2018, Racioppi2018, Karam2018a}, both in relation to gravity but also to beyond the Standard Model physics. The measured value of the scalar spectral index $n_s$, which is close to $1$, suggests a nearly scale-invariant power spectrum. But since exact scale invariance is excluded above $5\sigma$, the symmetry must be broken dynamically by quantum corrections via the Coleman--Weinberg mechanism~\cite{Coleman1973, Gildener1976}. Furthermore, by coupling the scalar field(s) to gravity in a non-minimal way, the Planck scale can by generated in a dynamical way through the vacuum expectation value (VEV) of the scalar(s). 

The paper is organised as follows. In the next section, we present the model in the Jordan frame. By using a Weyl transformation, we bring it to the Einstein frame, thus obtaining the tree-level potential which depends on two fields. Then, in Section~\ref{sec:GW}, we analyze the scalar potential by employing the Gildener-Weinberg formalism, which is a generalization of the Coleman-Weinberg mechanism for multiple fields. Ultimately we arrive at the one-loop corrected potential which is effectively one dimensional along the flat direction in the field space. In Section~\ref{sec:observables}, we use the one-loop potential to compute the inflationary observables and compare them to the experimental constraints. Finally, in Section~\ref{sec:conclusions} we summarize and conclude. 

\section{The Model}
\label{sec:model}

We start by presenting the model to be studied and defining our notations. In the context of classical scale invariance we consider the following Lagrangian density in the Jordan frame:
\be
\mathcal{L}^{J}=\sqrt{-\bar{g}} \left[\frac{\xi \phi^2 }{2}\bar{R}+\frac{\alp}{2} \bar{R}^2-\frac{1}{2} \bar{\nabla}^{\mu} \phi \bar{\nabla}_{\mu} \phi-\frac{\lambda_{\phi}}{4}\phi^4 \right]\,,
\label{JF-action}
\ee
where a bar indicates quantities in the Jordan frame. We have ommitted the Lagrangian ${\cal{L}}_m(\psi,\phi,\bar{g}_{\mu\nu})$ describing the rest of the matter fields interacting with the scalar $\phi$ and gravity through scale-invariant interactions, the details of which are not directly relevant to what we are about to discuss. It is known \cite{Dicke1962,Hwang1990,Hwang:1996np} that a general scalar-tensor theory of gravity $f(\phi,R)$ can be expressed in terms of an additional auxiliary scalar. In our case this amounts to writing the Lagrangian in the classically equivalent way
\be
\mathcal{L}^{J}=\sqrt{-\bar{g}} \left[\frac{1}{2}\left(\xi \phi^2+\alpha\chi^2\right)\bar{R}-\frac{\alpha}{8}\chi^4-\frac{1}{2} \bar{\nabla}^{\mu} \phi \bar{\nabla}_{\mu} \phi-\frac{\lambda_{\phi}}{4}\phi^4 \right]\,.
\label{JF-action-1}
\ee 
Then we consider a Weyl rescaling of the metric
\beq
g_{\mu \nu}= \Omega^2 \bar{g}_{\mu \nu}\,,{\label{WEYL}}
\eeq
where $g_{\mu \nu}$ is the Einstein frame metric and the conformal factor is
\be 
\Omega^2\,=\,\left(\alpha\chi^2+\xi\phi^2\right)/M_{\rm Pl}^2
\ee
or, equivalently expressed in terms of an auxiliary field $\zeta$ as 
\be 
\Omega^2=\frac{\zeta^2}{6M_{\rm Pl}^2} \ .
\ee
The scale $M_{\rm Pl}$ introduced by the conformal transformation \eqref{WEYL} will be identified with the Planck mass and will ultimately be related to VEV of the field $\zeta$.
The Lagrangian density in the Einstein frame takes the form
\beq
\mathcal{L}^{E}=\sqrt{-g}\left[ \frac{M_{\rm Pl}^2}{2} R-\frac{6 M_{\rm Pl}^2}{\zeta^2} \left(\frac{1}{2} \nabla^{\mu} \phi \nabla_{\mu} \phi+\frac{1}{2} \nabla^{\mu} \zeta \nabla_{\mu} \zeta \right)-V^{(0)E}(\phi, \zeta) \right] \,,
\label{EF-action}
\eeq
where the tree-level potential becomes
\be
V^{(0)E}(\phi, \zeta)= \frac{36 M_{\rm Pl}^4}{\zeta^4} \left[ \frac{\lambda_{\phi}}{4} \phi^4 + \frac{1}{8 \alp} \left( \frac{\zeta^2}{6}-\xi \phi^2 \right)^2 \right]\, .
\label{EF-potential}
\ee
It is convenient to rewrite \eqref{EF-potential} as
\be
V^{(0)E}=A \left(\frac{\phi}{\zeta}\right)^4+B \left(\frac{\phi}{\zeta}\right)^2+C \,,
\ee
where we have defined
\bea
A &\equiv &  9 M_{\rm Pl}^4\left(\lambda_{\phi}+\frac{\xi^2}{2\alpha}\right) \, ,\\
B &\equiv & -\frac{3\xi}{2\alpha} M_{\rm Pl}^4 \, ,\\
C &\equiv & \frac{M_{\rm Pl}^4}{8 \alp} \,.
\eea
The mass matrix for \eqref{EF-potential} can then be written in compact form as 
\beq
M_{ij}^2 \equiv \begin{pmatrix} 
\partial_{\phi}^2 V^{(0)E} & \partial_{\phi} \partial_{\zeta} V^{(0)E} \\
\partial_{\zeta} \partial_{\phi} V^{(0)E} & \partial_{\zeta}^2 V^{(0)E} 
\end{pmatrix}\biggr\rvert_{\phi = \upsilon_{\phi} , \zeta = \upsilon_{\zeta}} =
\frac{B}{3 M_{\rm Pl}^2}\begin{pmatrix} 
-2 & \sqrt{-\frac{2 B}{A}} \\
\sqrt{-\frac{2 B}{A}} & \frac{B}{ A} 
\end{pmatrix} \,,
\label{mass_matrix}
\eeq
where $\upsilon_{\phi}$ and $\upsilon_{\zeta}$ are the VEVs of the two fields. 


\section{Gildener--Weinberg Approach}
\label{sec:GW}

The tree-level potential~\eqref{EF-potential} contains two scalar fields which obtain nonzero VEVs, namely $\phi$ and $\zeta$. An elegant and simple formalism for analyzing spontaneous symmetry breaking due to quantum corrections in theories with multiple scalar fields was proposed by E. Gildener and S. Weinberg (GW)~\cite{Gildener1976}.

According to the GW approach~\cite{Gildener1976}, due to the running of the couplings of the theory, the tree-level potential is flat at some renormalization scale $\Lambda_{\rm GW}$ and has a degenerate valley of minima along a ray that extends out from the origin in field space. Then, by including the one-loop corrections, the effective potential obtains a radial shape along this ray and a non-zero VEV is dynamically generated, along with a mass for the radial component of the fields.

The conditions for the minimization of $V^{(0)E}$ yield the same constraint for the parameter space of the model, namely
\be\label{mins}
 \frac{\dd V^{(0)E}}{\dd \phi} \Big|_{\phi = v_\phi} = \frac{\dd \,V^{(0)E}}{\dd \zeta} \Big|_{\zeta = v_\zeta} = 0 \ ,
\ee
which gives
\be
 v_{\phi}^2 =- \frac{B}{2 A} v_{\zeta}^2 \,.
\label{constraint}
\ee
In the GW prescription, the mass matrix of the scalars $\phi$ and $\zeta$ contains the direction along the dynamically generated VEVs as one of its eigenvectors 
\be
\frac{1}{\sqrt{v^2_\phi + v^2_\zeta}}
\left(\begin{array}{c}
v_\phi \\
v_\zeta 
\end{array}
\right)\ .
\ee
The scalar masses may be diagonalized by a two-dimensional orthogonal rotation, parametrized by an angle $\omega$, as
\be
\left(\begin{array}{c}
\phi \\
\zeta 
\end{array}
\right)
=  
\left( \begin{array}{cc}
\cos\omega  & -\sin\omega   \\
\sin\omega   & \cos\omega 
\end{array}
\right)
\left(
\begin{array}{c}
s \\
\sigma 
\end{array}
\right),
\label{eq:rotation_matrix}
\ee
We can relate the VEVs through this mixing angle $\omega$ as
\be 
\tan\omega = \frac{v_\zeta}{v_\phi} \ . 
\label{eq:tanomega}
\ee
The diagonalization of the mass matrix \eqref{mass_matrix} then yields the following masses for the two scalar fields along the flat direction:
\be
m_{s}^2 = 0 \ ,
\ee
and 
\be
m_{\sigma}^2=\frac{\xi\left(\xi+12\lambda_{\phi} \alpha+6\xi^2\right)}{6\alpha\left(2\lambda_{\phi}\alpha+\xi^2\right)} M^2_{\rm Pl} \, ,
\ee
the first corresponding to the massless pseudo-Goldstone boson of broken scale invariance (scalon)
\be
s = \phi \cos\omega+\zeta\sin\omega\ ,
\label{s_definition}
\ee
while the second corresponds to the mass of the orthogonal state $\sigma=-\phi\sin\omega\,+\zeta\cos\omega$. 
Employing~\eqref{eq:tanomega}, along the flat direction the VEVs of the fields are related as
\be 
v_s = \frac{v_\phi}{\cos\omega} = \frac{v_\zeta}{\sin\omega} \ .
\ee
The kinetic terms of the original $\phi$ and $\zeta$ bosons are
\beq
\frac{6 M_{\rm Pl}^2}{\zeta^2} \left(\frac{1}{2} \nabla^{\mu} \phi \nabla_{\mu} \phi+\frac{1}{2} \nabla^{\mu} \zeta \nabla_{\mu} \zeta \right)=\frac{6 M_{\rm Pl}^2}{\zeta^2} \left( \frac{1}{2} \nabla^{\mu} s \nabla_{\mu} s+\frac{1}{2} \nabla^{\mu} s \nabla_{\mu} \sigma \right)\, ,
\eeq
where $\sigma=-\sin\omega\,\phi\,+\cos\omega\,\zeta$ is the orthogonal (massive) combination. Along the $\sigma=0$ direction in field space $\zeta\,=\,\sin\omega\,s\,+\cos\omega\,\sigma$ can be replaced with just $\zeta=\,s\,\sin\omega$ and its kinetic term reduced to
\be 
\frac{1}{2}\left(\frac{6 M_{\rm Pl}^2}{\sin^2\omega}\right)\frac{(\nabla s)^2}{s^2}\,=\,\frac{1}{2}\left(\nabla s_c\right)^2\,\,\Longrightarrow\,s_c - v_{s_c}\,=\,\frac{\sqrt{6}M_{\rm Pl}}{\sin\omega}\ln \frac{s}{v_s}\,
\ee
where $s_c$ is the corresponding canonical field. In the following, we only work with the canonically normalized inflaton $s$ and we omit the subscript $c$ for brevity.

At this point we may define the scale $M_{Pl}$ that was introduced by the conformal transformation in terms of the dynamically determined vev of the field $\zeta$ through the relation
\beq
v_{\zeta}^2 = 6 M_{\rm Pl}^2 \,, \qquad v^2 = \frac{6 M_{\rm Pl}^2}{\sin^2\omega} \ ,
\label{vevzeta}
\eeq 

The one-loop contribution to the potential, modulo derivative interactions of $\sigma$, in terms of the canonical field is given by \cite{Gildener1976, Farzinnia2016} 
\be
V^{(1)}=\frac{m^4_{\sigma}}{64 \pi^2 v_s^4} s^4\left[ \log \left(\frac{s^2}{v_s^2}\right)-\frac{1}{2} \right] \,,
\ee
where $v_s^2 = v_{\phi}^2+v_{\zeta}^2$ is the VEV of the pseudo-Goldstone boson of broken classical scale invariance $s$. We require the vanishing of the one-loop effective potential at the minimum. This requirement ensures that the cosmological constant is zero at the one-loop level. We have
\be 
V(v_s) \equiv  V^{(0)E}(v_s) + V^{(1)}(v_s) = 0 \ .
\ee
Then, the total one-loop effective potential along the flat direction is given by
\be
V(s)=\frac{m^4_{\sigma}}{128 \pi^2 }\left[ \frac{\sin^2\omega}{36 M_{\rm Pl}^4} s^4 \left(2 \ln \left[ \frac{s^2 \sin^2\omega}{6 M_{\rm Pl}^2} \right]-1 \right)+1  \right] \,.
\label{Potential final}
\ee
From the above potential we can obtain the radiatively generated mass for the $s$ boson
\be 
m_s^2 = \frac{\sin^2\omega}{48 \pi^2}\frac{m^4_\sigma}{M_{\rm Pl}^2}\ .
\ee
We see that the mass of $s$ is loop-suppressed with respect to that of $\sigma$. This means that the orthogonal state $\sigma$ effectively decouples during inflation and the pseudo-Goldstone boson $s$ plays the role of the inflaton along the flat direction.

Assuming a FLRW metric, the Klein--Gordon equation for the inflaton field $s$ has the form
\be 
\ddot{s} + 3 H \dot{s} + V'(s) = 0 \ ,
\label{eq:KG}
\ee
where $H = \dot{a}/a$ is the Hubble parameter and $a$ the scale factor.

\begin{figure}[h]
\centering
\includegraphics[width=0.7\textwidth]{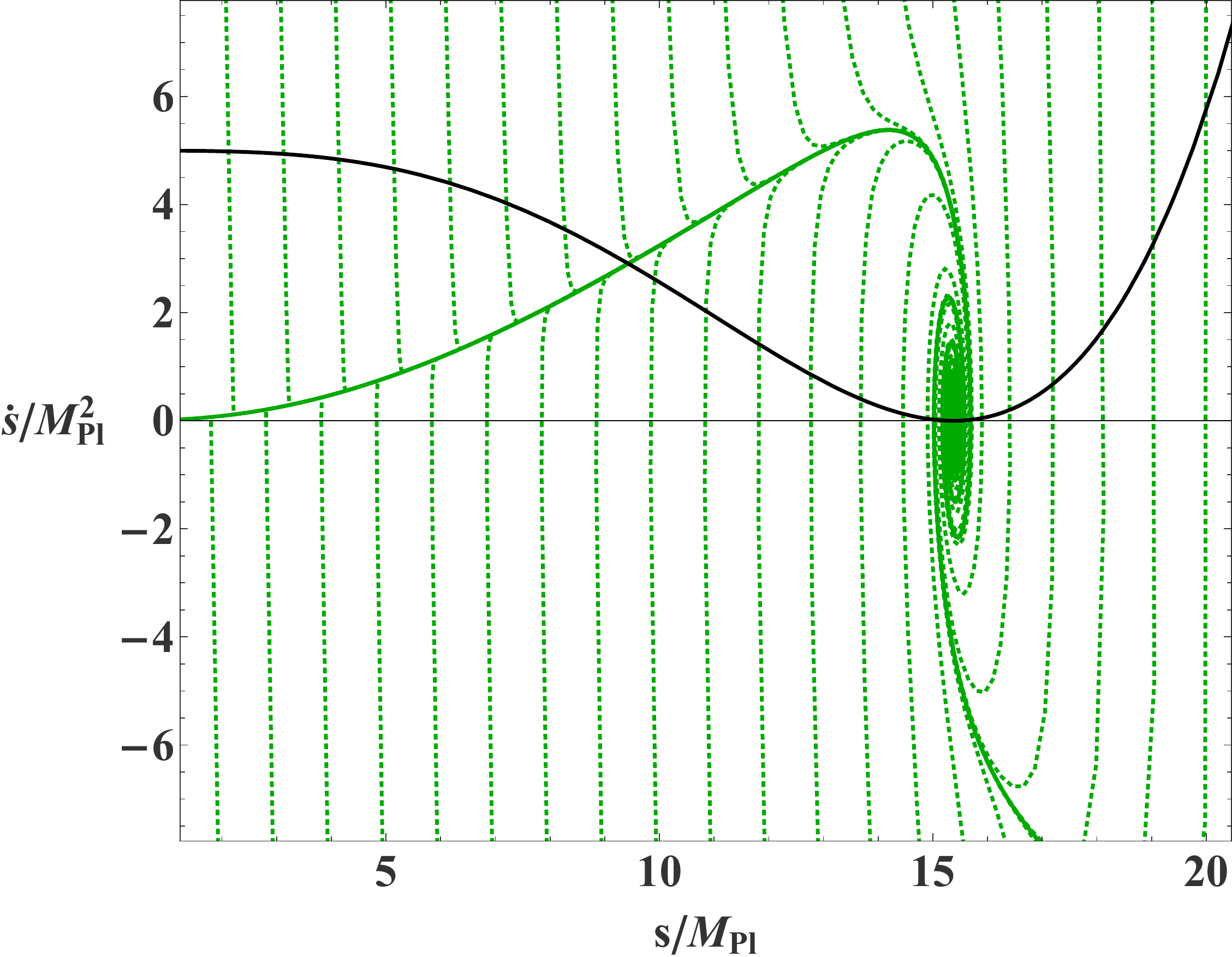}
\caption{\sf The attractor behavior of the one-loop effective potential~\eqref{Potential final} in the $s-\dot{s}$ phase space, where we have used $\xi = 0.3$, $\alpha = 0.01$ and $\omega = 0.16$. With the black curve we show the normalized ($5 V(s)/V(0)$) potential.}
\label{fig:attractor}
\end{figure}

In Fig.~\ref{fig:attractor} we fix the values of the parameters to $\xi = 0.3$, $\alpha = 0.01$ and $\omega = 0.16$, we numerically solve the Klein--Gordon equation~\eqref{eq:KG} for a plethora of initial conditions for the inflaton $s$ and plot the $s-\dot{s}$ phase space. Clearly, the potential exhibits an attractor behavior since, regardless of the initial conditions, all trajectories in the phase space (green-dotted curves) quickly converge in a single trajectory that terminates at the location of the minimum.

Notice that the conditions \eqref{constraint} and \eqref{vevzeta} constrain the free parameters of the model. As a result we find that
\beq
v_{\phi}^2=v^2-v_{\zeta}^2 =6 M_{\rm Pl}^2 \cot^2\omega\,.
\label{vevphi}
\eeq
Then, upon substituting \eqref{vevzeta} and \eqref{vevphi} into \eqref{constraint} one obtains
\beq
2 A \cot^2\omega=-B \,.
\eeq
Equivalently we have the constraint
\beq
\left(12 \alp \lambda_{\phi}+6 \xi^2 \right)\cot^2\omega=\xi \,,
\label{final constraint}
\eeq
and so we are left with 3 free parameters. If we choose to solve \eqref{final constraint} for the coupling $\lambda_{\phi}$ we have that\footnote{Note that we have ignored the running of the couplings $\xi$ and $\alpha$, assuming that their $\beta$-functions are such that their values do not vary much over the inflaton field excursion.}
\beq
\lambda_{\phi}=\frac{\xi}{12 \alp} \left[ \tan^2\omega-6 \xi \right] \,.
\eeq


\section{Observables}
\label{sec:observables}

The imposition of the constraint \eqref{final constraint}, has reduced the number of the free parameters of the system to three, namely the non-minimal coupling of the field ($\xi$), the Starobinsky parameter ($\alp$), and the mixing angle ($\omega$). As we shall now demonstrate, it is only the angle $\omega$ that affects the predictions for  observables and the other two parameters come into play via the VEVs that determine the value of $\omega$.

In the slow-roll approximation, the expressions for the tensor-to-scalar ratio $r$, the tilt $n_s$ of the scalar power spectrum and the running $\alpha_s$ of the scalar spectral index are given in terms of the values of the potential slow-roll parameters $\eps_V$, $\eta_V$ and $\xi_V^{2}$ at horizon crossing as
\beq
r \approx 16 \eps_V^* \,,
\label{r}
\eeq
\beq
n_s \approx 1- 6 \eps_V^* + 2 \eta_V^* \,,
\label{ns}
\eeq
and
\be 
\alpha_s = 16 \eps_V^* \eta_V^* -24 \eps_V^{2*} - 2 \xi_V^{2*} \ ,
\label{alphas}
\ee
where
\beq
\eps_V \equiv \frac{M_{\rm Pl}^2}{2} \left( \frac{ V'(s)}{V(s)} \right)^2 \ , \quad \eta_V \equiv M_{\rm Pl}^2 \frac{ V''(s)}{V(s)}\,, \quad \xi^2_V \equiv M^4_{\rm Pl} \frac{V'(s)V'''(s)}{V(s)^2} \ ,
\eeq
and $^*$ has been used to denote the value of the slow-roll parameters at horizon crossing.

By performing a direct computation it is straightforward to show that the slow-roll parameters for the potential \eqref{Potential final} take the form
\beq
\eps_V = M_{\rm Pl}^2 \frac{32 s^6\,  \log{\left[ \frac{6 M_{\rm Pl}^2}{s^2 \sin^2\omega} \right]} \sin^8\omega }{\left[ 36 M_{\rm Pl}^4 + s^4 \sin^4\omega \left(2 \log{\left[ \frac{s^2 \sin^2\omega}{6 M_{\rm Pl}^2}  \right]} - 1 \right)  \right]^2}  \ ,
\label{epsilonV}
\eeq
\beq
 \eta_V = M_{\rm Pl}^2 \frac{8 s^2 \left[2+3 \log{\left(\frac{s^2 \sin^2\omega}{6 M_{\rm Pl}^2} \right)}  \right] \sin^4\omega }{36 M_{\rm Pl}^4 + s^4 \sin^4\omega \left(2\log{\frac{s^2 \sin^2\omega}{6 M_{\rm Pl}^2 }}-1 \right)} \ ,
 \label{etaV}
\eeq
and
\beq
\xi_V^{2} = M_{\rm Pl}^4  \frac{128 s^4 \sin^8\omega  \log \left(\frac{s^2 \sin^2\omega}{6 M_{\rm Pl}^2}\right) \left(3 \log \left(\frac{s^2 \sin^2\omega}{6 M_{\rm Pl}^2}\right)+5\right)}{\left(s^4 \sin ^4\omega \left(2 \log \left(\frac{s^2 \sin ^2\omega}{6 M_{\rm Pl}^2}\right)-1\right)+36 M_{\rm Pl}^4\right)^2} \ ,
 \label{xiV}
\eeq
Thus it is evident that the sole parameter that plays a role in the expressions \eqref{r} and \eqref{ns} for the observables is the mixing angle $\omega$\footnote{This is already clear from \eqref{Potential final} where the parameters $\xi$ and $\alp$ enter the expression of the potential only via the mass $m_\zeta$ and the latter cancels out in the expressions for $\eps_V$, $\eta_V$ and $\xi_V^{2}$.}. Note that an inflaton field excursion can occur on either side of the minimum of the potential~\eqref{Potential final}, corresponding to small and large field inflation. In the following, however, we will not deal with the large field scenario since it yields a tensor-to-scalar ratio which is excluded by observations.

\begin{figure}[h]
\centering
\includegraphics[width=0.7\textwidth]{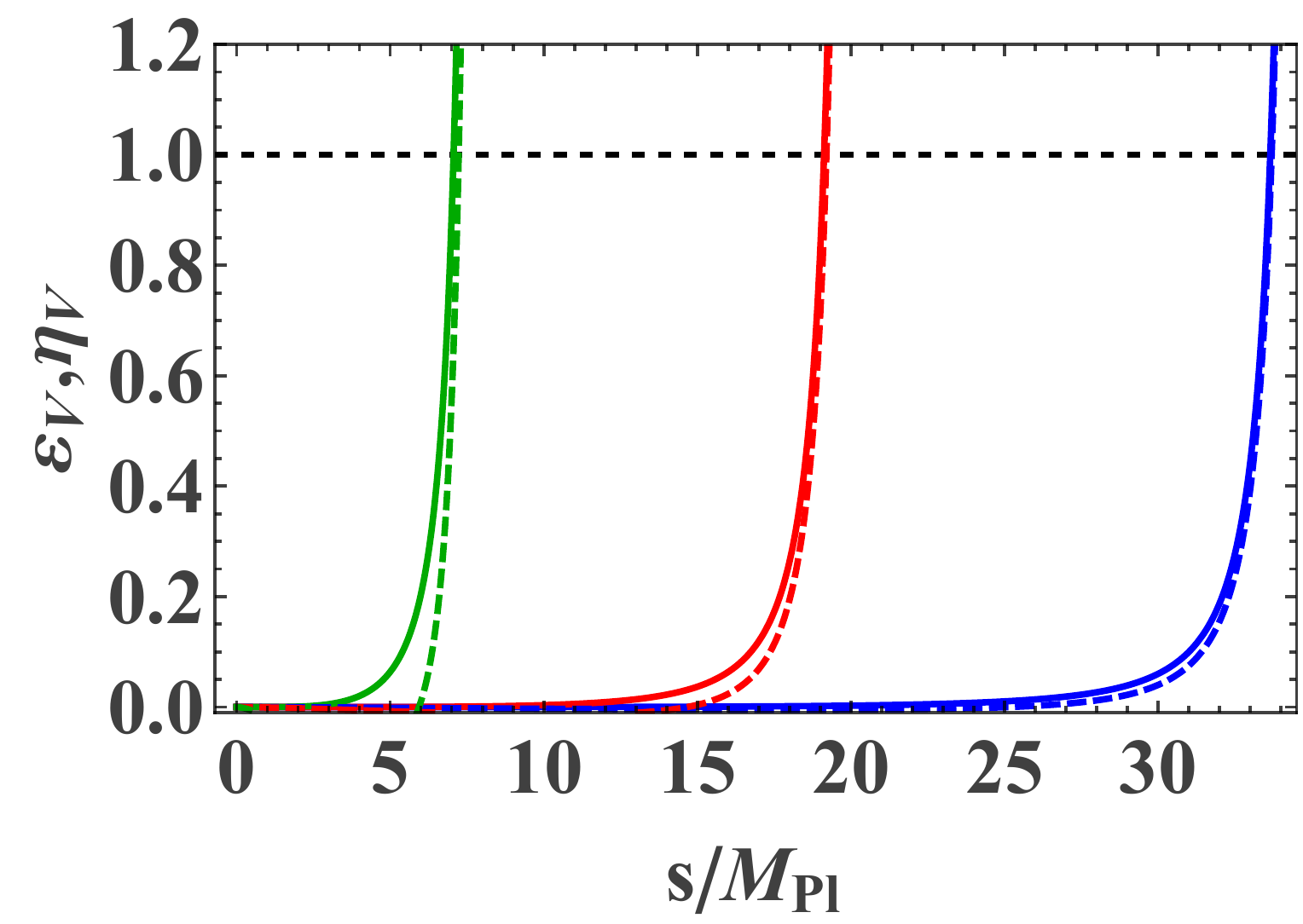}
\caption{\sf The slow-roll parameters $\epsilon_V$ (solid curves) and $\eta_V$ (dashed curves) as a function of the inflaton field $s$ for fixed values of the mixing angle at $\sin\omega = 0.07$ (blue curves), $\sin\omega = 0.12$ (red curves) and $\sin\omega = 0.30$ (green curves).}
\label{fig:PSRPs}
\end{figure}

In Fig.~\ref{fig:PSRPs}, we plot the slow-roll parameters $\epsilon_V$ and $\eta_V$, given in Eqs.~\eqref{epsilonV}--\eqref{etaV}, as a function of the inflaton field $s$ for a few fixed values of the mixing angle $\omega$. We observe that both slow-roll parameters become unity around the same inflaton field value.

The number $N$ of e-folds of inflation elapsed in the Einstein frame is defined as $\dd N=H\dd t$ where $H$ and $t$ are the Hubble parameter and time coordinate respectively. In terms of the slow roll parameter $\eps_V$ one has
\beq
N = \int _{s_*}^{s_f} \frac{1}{\sqrt{2 \eps_V(s)}} \dd s\,,
\eeq
where $s_*$ is the value of the inflaton field at the time of horizon crossing while $s_f$ is its value at the end of inflation.

In the case of the model under consideration, the total number of e-folds is given by the following analytic expression:
\be
N 
= \left\lbrace 
\frac{3}{8 \sin^2\omega} 
\left[
{\rm li}\left( \frac{6 \sin^2 \omega\, M_{\rm Pl}^2}{s^2}\right)
-{\rm li}\left(\frac{\sin^2 \omega\,s^2}{6 M_{\rm Pl}^2}\right)
\right]
+\frac{s^2}{8 M_{\rm Pl}^2}\right\rbrace  \Bigg|_{s_{f}}^{s_{*}} \ ,
\ee
where 
${\rm li}(z)$ is the logarithmic integral (${\rm li}(z)=\int_{0}^{z}\frac{1}{\log t} dt$).

The required amount of inflation in order for the horizon and flatness problems to be solved is $N \approx 60$. Inflation ends exactly when the first Hubble slow-roll parameter $\eps_H \equiv -\dot{H} / H^{2}=1$ but in the slow-roll approximation $\eps_H \approx \eps_V$ and so it is often the condition $\eps_V = 1$ that is used instead in order to obtain $s_f$.
 
\begin{figure}[h]
\centering
\includegraphics[width=0.7\textwidth]{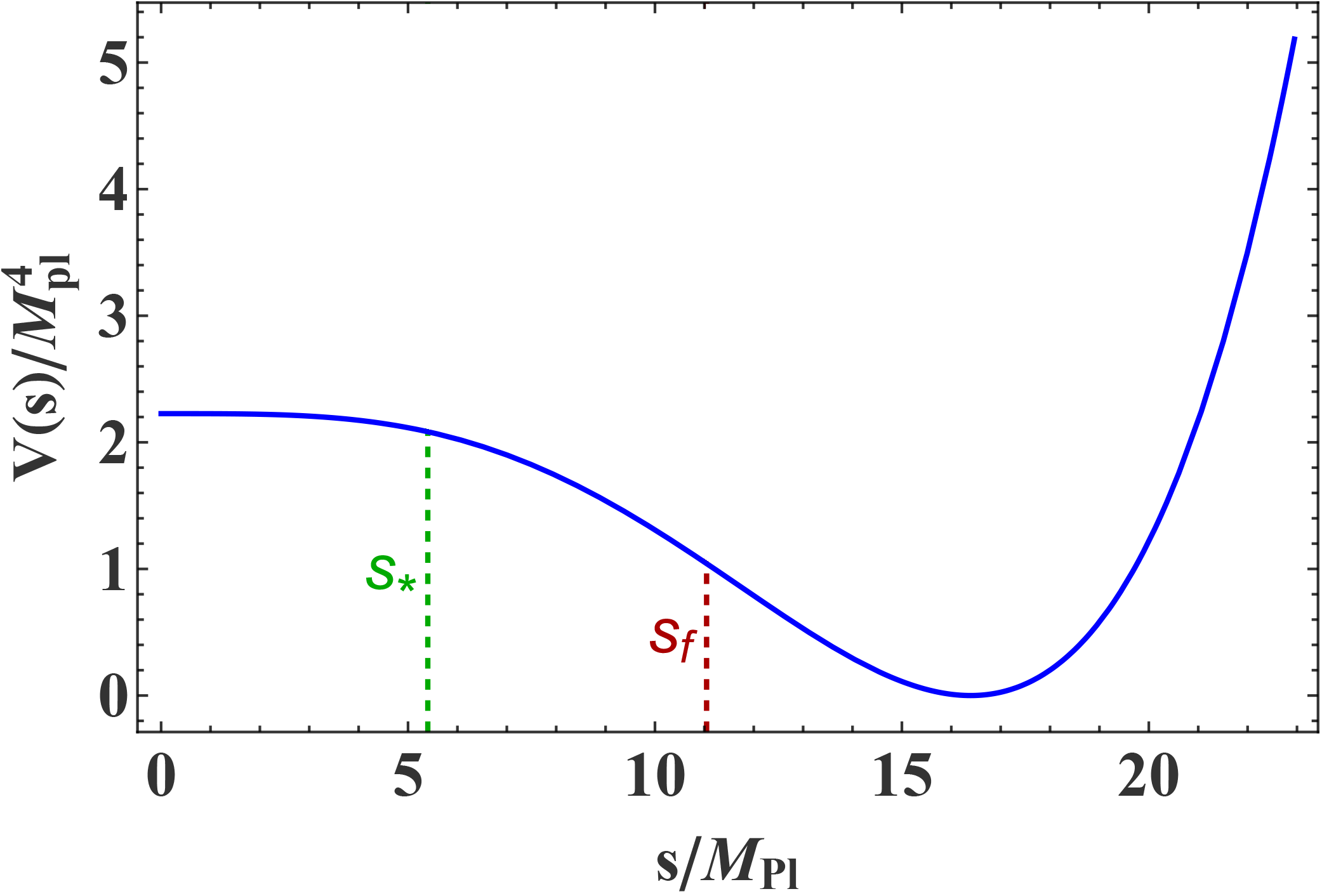}
\caption{\sf An inflaton excursion that yields $N=59.3$ when $\sin\omega= 0.15$. Inflation starts(ends) at the green(red)-dashed line. The values of the observables in this case are $(r,n_s)=(0.014,0.963)$.}
\label{fig:ExcursionExample}
\end{figure}

An example of an inflaton excursion in our model \eqref{Potential final} that yields a sufficient amount of inflation is given in Fig.\ref{fig:ExcursionExample}. The red-dashed line corresponds to the value $s_f$ of the inflaton where inflation ends subject to the condition $\eps_V=1$. The green-dashed line corresponds to the value $s_*$ of the inflaton at the time of horizon crossing that is obtained by the requirement $N \approx 60$.
By considering different values for $s_*$ we vary the total number of e-folds of inflation elapsed during the field excursion and consequently obtain different predictions for the observables $r$ and $n_s$.

\begin{figure}[h]
\centering
\includegraphics[width=0.7\textwidth]{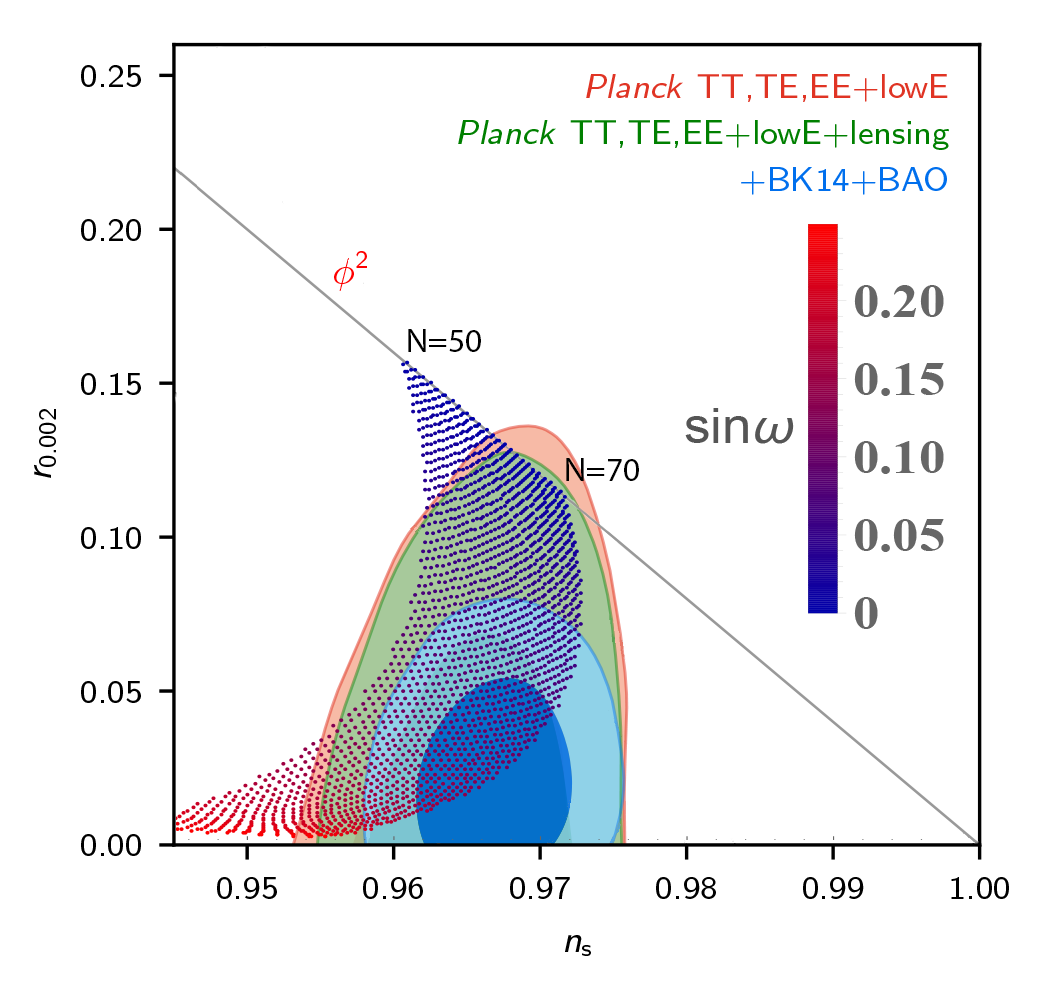}
\caption{\sf The predictions of the model \eqref{Potential final} for a wide range of values for the mixing angle $\omega$ and for field excursions that yield sufficient amount of inflation i.e. 50 to 70 e-folds.}
\label{fig:rnsFinalplot}
\end{figure}

In Fig.\ref{fig:rnsFinalplot} we considered different values for the mixing angle ranging from $\sin\omega = 0$ up to $\sin\omega = 0.23 $ that yield $50\leqslant   N \leqslant 70$ and plot the predictions of the model in the $n_s-r$ plane against the current bounds set by the Planck collaboration~\cite{Akrami2018}. One can see that for small values of the mixing angle the predictions of the model converge to the ones obtained by the minimal quadratic inflation model. This is explained by the fact that the potential~\eqref{Potential final}, for $\sin\omega \rightarrow 0$ and around its minimum, is approximated by
\be 
V(s) \simeq \frac{m^4_\sigma \sin^2\omega}{96\pi^2 M^2_{\rm Pl}} \left( s - v_s  \right)^2 \ .
\ee 

\begin{figure}[H]
\centering
\includegraphics[width=0.6\textwidth]{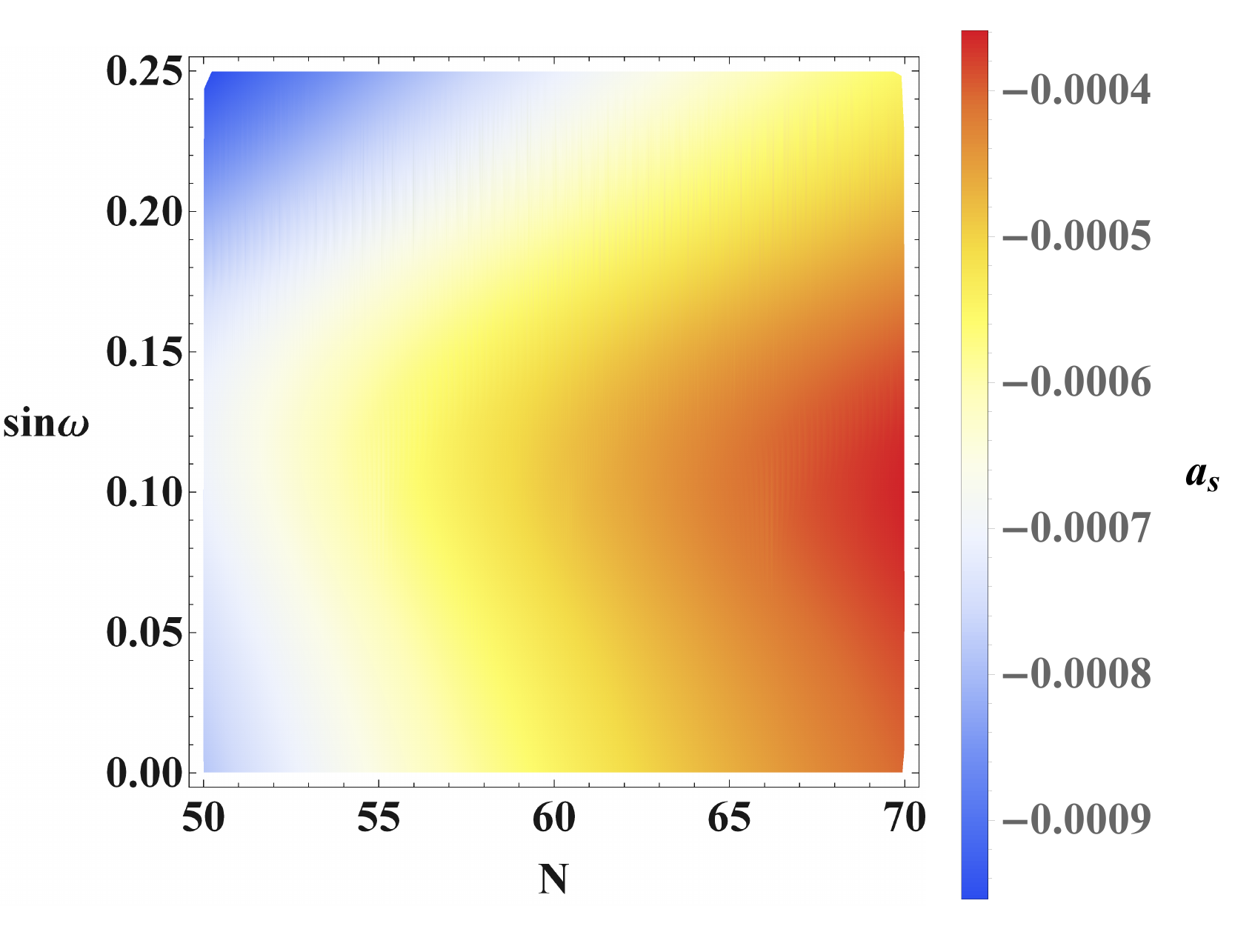}
\caption{\sf The running of the scalar spectral index $\alpha_s$ for values of the mixing angle in the range $\sin\omega = 0-0.25$ and for a number of e-folds between $N=50-70$.} 
\label{fig:running}
\end{figure}

Finally, in Fig.~\ref{fig:running}, for various values of the mixing angle $\omega$ and for $N=50-70$ e-folds we plot the running of the scalar spectral index $\alpha_s$. All of the resulting values are compatible with the observational constraints~\cite{Akrami2018} $\alpha_s = -0.0045 \pm 0.0067$ at $68\%$ CL.


\section{Summary and Conclusions}
\label{sec:conclusions}

In this work, we considered slow-roll inflation in the theory of a scalar field with scale-imvariant interactions coupled to gravity non-minimally in the presence of an $R^2$ term. The gravitational sector in the Jordan frame consists of a nonminimal coupling term of the inflaton to the Ricci scalar and an $R^2$ correction. Expressing the action in terms of an auxiliary scalar field and  transforming it to the Einstein frame by means of a Weyl rescaling of the metric, the theory consists of two real scalar fields with non-canonical kinetic terms subject to a two-dimensional inflationary potential. By applying the formalism of Gildener and Weinberg we were able to effectively describe inflation by means of a single scalar field ($s$), the pseudo-Goldstone boson along the flat direction of the tree-level potential. We obtained the one-loop corrected potential along the flat direction and in turn, by numerically solving the Klein-Gordon equation, we have verified that it exhibits an inflationary attractor behavior. 

Then, we studied the predictions for the tensor-to-scalar ratio ($r$), the scalar spectral index ($n_s$) and the running of the scalar spectral index ($\alpha_s$). We found that the only free parameter that affects the predictions is the mixing angle ($\omega$) between the two scalar degrees of freedom. For various field excursions of the inflaton $s$ that produce a sufficient amount of inflation, i.e. $50$ to $70$ e-folds, and for a wide range of values for $\omega$ we found that the model yields predictions that lie well within the recent, strict bounds for the allowed region in the ($n_s-r$) parametric space set by the Planck collaboration. For the aforementioned field excursions and values of the mixing angle the predictions for the running of the spectral index are also in excellent agreement with the observations.

In conclusion, the present model is able to successfully describe inflation and at the same time generate the Planck scale in a dynamical way by means of the Coleman-Weinberg mechanism. The inclusion of the $R^2$ term affects mainly the predictions of $r$. Larger values of the Starobinsky coupling $\alpha$ correspond to larger values of the mixing angle $\omega$, which in turn result to smaller values for $r$.

\section*{Acknowledgements}

This work is implemented through the Operational Program ``Human Resources Development, Education and Lifelong Learning" and is co-financed by the European Union (European Social Fund) and Greek national funds.

\bibliography{References}{}
\bibliographystyle{utphys}

\end{document}